\documentclass[10pt,conference]{IEEEtran}
\IEEEoverridecommandlockouts

\usepackage{cite}
\usepackage{amsmath,amssymb,amsfonts}
\usepackage{algorithmic}
\usepackage{graphicx}
\usepackage{textcomp}
\usepackage{xcolor}
\usepackage{xspace} 
\usepackage{listings}
\usepackage{hyperref}
\usepackage{enumitem}
\usepackage{subcaption} 

\setlist{leftmargin=15pt,labelindent=15pt}
\setlist[enumerate]{wide=0pt, widest=99,leftmargin=\parindent, labelsep=*}

\definecolor{codegreen}{rgb}{0,0.6,0}
\definecolor{codegray}{rgb}{0.5,0.5,0.5}
\definecolor{codepurple}{rgb}{0.58,0,0.82}
\definecolor{backcolour}{rgb}{0.95,0.95,0.92}
\lstdefinestyle{mystyle}{
    backgroundcolor=\color{backcolour},   
    commentstyle=\color{codegreen},
    keywordstyle=\color{magenta},
    numberstyle=\tiny\color{codegray},
    stringstyle=\color{codepurple},
    basicstyle=\ttfamily\footnotesize,
    breakatwhitespace=false,         
    breaklines=true,                 
    captionpos=b,                    
    keepspaces=true,                 
    numbers=left,                    
    numbersep=5pt,                  
    showspaces=false,                
    showstringspaces=false,
    showtabs=false,                  
    tabsize=1
}

\definecolor{verylightgray}{rgb}{.97,.97,.97}

\lstdefinelanguage{Solidity}{
	keywords=[1]{address, anonymous, assembly, assert, balance, break, call, callcode, case, catch, class, constant, continue, constructor, contract, debugger, default, delegatecall, delete, do, else, emit, event, experimental, export, external, false, finally, for, function, gas, if, implements, import, in, indexed, instanceof, interface, internal, is, length, library, log0, log1, log2, log3, log4, memory, modifier, msg, new, payable, pragma, private, protected, public, pure, push, require, return, returns, revert, selfdestruct, send, sender, solidity, storage, struct, suicide, super, switch, then, this, throw, transfer, true, try, typeof, using, value, view, while, with, addmod, ecrecover, keccak256, mulmod, ripemd160, sha256, sha3}, 
	keywordstyle=[1]\color{blue}\bfseries,
	keywords=[2]{bool, byte, bytes, bytes1, bytes2, bytes3, bytes4, bytes5, bytes6, bytes7, bytes8, bytes9, bytes10, bytes11, bytes12, bytes13, bytes14, bytes15, bytes16, bytes17, bytes18, bytes19, bytes20, bytes21, bytes22, bytes23, bytes24, bytes25, bytes26, bytes27, bytes28, bytes29, bytes30, bytes31, bytes32, enum, int, int8, int16, int24, int32, int40, int48, int56, int64, int72, int80, int88, int96, int104, int112, int120, int128, int136, int144, int152, int160, int168, int176, int184, int192, int200, int208, int216, int224, int232, int240, int248, int256, mapping, string, uint, uint8, uint16, uint24, uint32, uint40, uint48, uint56, uint64, uint72, uint80, uint88, uint96, uint104, uint112, uint120, uint128, uint136, uint144, uint152, uint160, uint168, uint176, uint184, uint192, uint200, uint208, uint216, uint224, uint232, uint240, uint248, uint256, var, void, ether, finney, szabo, wei, days, hours, minutes, seconds, weeks, years},	
	keywordstyle=[2]\color{teal}\bfseries,
	keywords=[3]{block, blockhash, coinbase, difficulty, gaslimit, number, timestamp, , data, gas, sig, value, now, tx, gasprice, origin},	
	keywordstyle=[3]\color{violet}\bfseries,
	identifierstyle=\color{black},
	sensitive=false,
	comment=[l]{//},
	morecomment=[s]{/*}{*/},
	commentstyle=\color{gray}\ttfamily,
	stringstyle=\color{red}\ttfamily,
	morestring=[b]',
	morestring=[b]"
}

\def\BibTeX{{\rm B\kern-.05em{\sc i\kern-.025em b}\kern-.08em
    T\kern-.1667em\lower.7ex\hbox{E}\kern-.125emX}}

\begin{document}

\title{\mytool: Smart Contracts Reentrancy and Unhandled Exceptions Vulnerability Dataset\\
}







\author{
\IEEEauthorblockN{
    Chavhan Sujeet Yashavant\textsuperscript{1}, 
    MitrajSinh Chavda\textsuperscript{1}, 
    Saurabh Kumar\textsuperscript{2}, 
    Amey Karkare\textsuperscript{1}, 
    Angshuman Karmakar\textsuperscript{1}}
\IEEEauthorblockA{
    \textsuperscript{1}\textit{Department of Computer Science and Engineering}, 
    \textit{Indian Institute of Technology Kanpur}, Kanpur, India \\
    Email: \{sujeetc, mbchavda, karkare, angshuman\}@cse.iitk.ac.in}
\IEEEauthorblockA{
    \textsuperscript{2}\textit{Department of Computer Science and Engineering}, 
    \textit{Indian Institute of Technology Hyderabad}, Hyderabad, India \\
    Email: saurabhkr@cse.iith.ac.in}
}
\newcommand{\mytool}{{\scshape SCRUBD\xspace}}

\newcommand{\mytoolcd}{{\scshape \mytool/{\scalebox{0.8}{CD}}\xspace}}

\newcommand{\mytoolsd}{{\scshape \mytool/{\scalebox{0.8}{SD}}\xspace}}

\newcommand{\tbd}[1]{{\color{red}\bf TBD: #1}}

\newcommand{\crossmark}{\ensuremath{\boldsymbol{\times}}}

\newcommand{\totalLabelledRENTSCinSmartbugs}{{31}}
\newcommand{\totalLabelledRENTFuninSmartbugs}{{177}} 
\newcommand{\totalLabelledBuggyFuninSmartbugs}{{31}} 
\newcommand{\totalLabelledBuggyUniqueFuninSmartbugs}{{46}} 
\newcommand{\totalUnlabelledSCinSmartbugs}{{47,518}}

\newcommand{\uniqueContractsInMengRen}{{??}}


\newcommand{\totalCollectedSCinTTR}{{230,548}}
\newcommand{\totalUniqueSCinTTR}{{139,424}}
\newcommand{\totalToolFlagSCinTTR}{{21,212}}
\newcommand{\totalToolFlagTrueSCinTTR}{{41}} 
\newcommand{\totalToolFlagTrueUniqueFuninTTR}{{20}}

\newcommand{\reentCall}{{reent-call}}

\newcommand{\condCall}{{conditional call()}}

\newcommand{\totalSCtxGTOne}{{43,83,982}}

\newcommand{\totalSCtxGTOnebalGTZero}{{3,96,471}}

\newcommand{\extractedSCtxGTOnebalGTZero}{{3,57,664}}

\newcommand{\extractedSCtxGTOne}{{21,58,515}}

\newcommand{\uniqueExtractedSCtxGTOne}{{4,54,895}}

\newcommand{\uniqueExtractedSCtxGTOnebalGTZero}{{37,814}}

\newcommand{\totalSCInCrowd}{{469}}

\newcommand{\totalFunsMarkedVulnByTools}{2080}
\newcommand{\totalFunsinCrowdsource}{{746}}

\newcommand{\totalRENTFunsInCrowd}{{746}}
\newcommand{\totalRENTStudentFunsInCrowd}{{566}}
\newcommand{\totalRENTNonStudentFunsInCrowd}{{180}}

\newcommand{\totalRENTVulnFunsInCrowd}{{245}}
\newcommand{\totalRENTNonVulnFunsInCrowd}{{501}}

\newcommand{\totalStudentsinCrowdsource}{{63}}

\newcommand{\totalUEFunsInCrowd}{{566}}
\newcommand{\totalUEVulnFunsInCrowd}{{275}}
\newcommand{\totalUENonVulnFunsInCrowd}{{291}}

\newcommand{\totalFunsinManualDB}{{239}}

\newcommand{\totalRENTVulnFunsInManual}{{155}}

\newcommand{\totalRENTNonVulnFunsInManual}{{84}}

\newcommand{\totalRENTFunsInScrawlD}{{985}} 

\newcommand{\uniqueSCtxGTOne}{{4,10,995} }

\newcommand{\uniqueSCtxGTOneCallInst}{{3,10,995} }

\newcommand{\uniqueSCtxGTOneCallInstFunctionDd}{{50,000} }

\newcommand{\uniqueSCtxGTOneCallInstFunctionDdMultiCall}{{5,000} }

\newcommand{\uniqueSCtxGTOneCallInstFunctionDdSingleCall}{{45,000} }

\maketitle

\begin{abstract}
    

Smart Contracts (SCs) handle transactions in the Ethereum blockchain worth millions of United States dollars, making them a lucrative target for attackers seeking to exploit vulnerabilities and steal funds. The Ethereum community has developed a rich set of tools to detect vulnerabilities in SCs, including reentrancy (RE) and unhandled exceptions (UX). A dataset of SCs labelled with vulnerabilities is needed to evaluate the tools' efficacy. Existing SC datasets with labelled vulnerabilities have limitations, such as covering only a limited range of vulnerability scenarios and containing incorrect labels. As a result, there is a lack of a standardized dataset to compare the performances of these tools. \mytool~aims to fill this gap. We present a dataset of real-world SCs and synthesized SCs labelled with RE and UX. The real-world SC dataset is labelled through crowdsourcing, followed by manual inspection by an expert, and covers both RE and UX vulnerabilities. On the other hand, the synthesized dataset is carefully crafted to cover various RE scenarios only. Using~\mytool~we compared the performance of six popular vulnerability detection tools. Based on our study, we found that Slither outperforms other tools on a crowdsourced dataset in detecting RE vulnerabilities, while Sailfish outperforms other tools on a manually synthesized dataset for detecting RE. For UX vulnerabilities, Slither outperforms all other tools.




\end{abstract}

\section{Introduction}

Public blockchains like Ethereum host millions of smart contracts (SCs) that can be invoked by anyone via public functions~\cite{Smart-Contracts-and-Opportunities-for-Formal-Methods}.  SCs manage Ether, Ethereum's cryptocurrency, which is valued in various fiat currencies like USD (United States Dollars), EUR (Euro), and INR (Indian Rupees). Hence, SCs are crucial for Ethereum. However, their significance makes them lucrative targets for malicious actors seeking financial gain or intending to cause disruption. Consequently, attacks on Ethereum SCs that target vulnerabilities in SCs have sharply increased over recent years~\cite{A-Survey-on-Ethereum-Systems-Security-Vulnerabilities-Attacks-and-Defenses, A-Survey-of-Attacks-on-Ethereum-Smart-Contracts, scattack1, scattack2, scattack3}. This increase in attacks highlights the urgent need for stronger security measures and robust protocols in the Ethereum ecosystem.

To  strengthen the security of SCs, the blockchain community has developed a diverse array of tools to detect vulnerabilities in SCs~\cite{slither,oyente,securify,smartcheck,mythril,maian,sailfish}. Testing using a quality test suite (dataset) is integral to any tool development process. The developers use these datasets to assess efficacy of vulnerability detection tools. However, it is difficult to compare these metrics for different tools as each tool relies on a different dataset~\cite{Empirical-Evaluation-of-Smart-Contract-Testing:What-is-the-Best-Choice, slither, smartcheck, sailfish}. For a fair and consistent comparison of vulnerability detection tools, there is a pressing need to create a high-quality sound dataset. Further, a new vulnerability detection tool can be tested on this dataset and refined progressively.

We outline the limitations of some well-known SC vulnerability datasets in Section~\ref{sec:limitations}. To overcome the limitations of existing SC vulnerability datasets, we present~\mytool. In this work, we focus on RE and UX vulnerabilities, as an earlier study~\cite{usenix-exploitation} has shown that they are the most commonly exploited vulnerabilities. Contributions of this work are as follows:

\begin{table}[t]
\caption{Total Vulnerable (V) and Non-Vulnerable (NV) Functions in \mytoolcd~and \mytoolsd}
\centering
\resizebox{0.5\textwidth}{!}{  
\begin{tabular}{|l|c|c|c|}
\hline
\textbf{Dataset} & \textbf{Methodology} & \textbf{RE} & \textbf{UX} \\ \hline
\mytoolcd & Crowdsourcing  & \totalRENTVulnFunsInCrowd~V, \totalRENTNonVulnFunsInCrowd~NV & \totalUEVulnFunsInCrowd~V, \totalUENonVulnFunsInCrowd~NV \\ \hline
\mytoolsd & Synthesized Dataset & \totalRENTVulnFunsInManual~V, \totalRENTNonVulnFunsInManual~NV & N/A \\ \hline
\end{tabular}
}
\label{table:vulnerability-overview}
\end{table}
\begin{enumerate}
    \item 
        \mytool~consists of two datasets: \mytoolcd~(crowdsourced dataset) and \mytoolsd~(synthesized dataset).
        \begin{enumerate}
        \item \mytoolcd: We developed a website that lists SC functions and asked users to flag them as vulnerable or non-vulnerable. The website lists a function in a SC only if it is flagged as vulnerable by at least one of the following tools: Conkas~\cite{conkas}, Mythril~\cite{mythril}, Sailfish~\cite{sailfish}, Slither~\cite{slither}, Solhint~\cite{solhint}. This was assigned as a course assignment in a graduate-level course. An experienced researcher manually reviewed and validated the students' answers. The resulting crowdsource dataset includes~\totalSCInCrowd~contracts. Within~\totalSCInCrowd~contracts,~\totalFunsinCrowdsource~functions are labelled for RE vulnerability and~\totalUEFunsInCrowd~functions are labelled for UX vulnerability (see Table~\ref{table:vulnerability-overview}).
        \item \mytoolsd: We have used a novel approach to manually craft a dataset for RE vulnerability. We have analyzed various scenarios that lead to reentrancy (RE) vulnerabilities and created a manually synthesized dataset, which includes test cases for both vulnerable and non-vulnerable scenarios. A minor modification in a test case can shift it from being vulnerable to non-vulnerable and vice versa.~\mytoolsd~contains a total of~\totalFunsinManualDB~cases, out of which~\totalRENTVulnFunsInManual~are vulnerable and~\totalRENTNonVulnFunsInManual~are non-vulnerable (see Table~\ref{table:vulnerability-overview}). 
    \end{enumerate} 
 
	\item We compared the performance of six SC vulnerability analysis tools using~\mytool. We present an evaluation showing Slither outperforms other tools in detecting RE vulnerabilities on~\mytoolcd, while Sailfish excels on~\mytoolsd, and Slither leads in detecting UX vulnerabilities (see Section~\ref{sec:results}). Further, we found anomalies in an existing dataset, Turn-The-Rudder~\cite{turn-the-rudder-icse}. We identified misleading labels in the dataset, a finding that has been confirmed by the authors (see Section~\ref{sec:limitations}).
 

    \item We have made our dataset and website publicly available\footnote{\href{https://github.com/sujeetc/SCRUBD}{https://github.com/sujeetc/SCRUBD},~\href{https://scaudit.cse.iitk.ac.in}{https://scaudit.cse.iitk.ac.in}}. Researchers can use~\mytool~to conduct unbiased evaluations of both new and existing tools. 
\end{enumerate}

\section{Limitations of Existing RE and UX Datasets}
\label{sec:limitations}

This section outlines the limitations of existing datasets, and we have made these limitations publicly available\footnote{\href{https://github.com/sujeetc/SCRUBD/tree/main/limitations}{https://github.com/sujeetc/SCRUBD/tree/main/limitations}}.

\subsection{Turn-The-Rudder Dataset}
\label{limit:ttr}

Zheng et al.\cite{turn-the-rudder-icse} used crowdsourcing to annotate smart contracts (SCs) for RE. After crowdousrcing, they discovered that~\totalToolFlagTrueSCinTTR~contracts have RE out of~\totalToolFlagSCinTTR, with each contract having a vulnerable function. However, when we manually reviewed the non-vulnerable SCs, we found that they contained the exact copy of a few functions from the vulnerable contracts. We conveyed our findings to authors with five such entries, and they confirmed our findings and updated their GitHub\footnote{\href{https://github.com/InPlusLab/ReentrancyStudy-Data/commit/5b6be0f6be24591f1455e3467d956c397357abd3}{GitHub-commit-link}}. So there are \textit{46} vulnerable functions now. We applied function-level de-duplication and identified~\totalToolFlagTrueUniqueFuninTTR~unique functions with reentrancy. Upon semantically matching these 20 functions, we found that only 14 unique functions remained. We found that the non-vulnerable section contains 71 functions that match exactly with vulnerable functions flagged by the dataset and therefore are potentially vulnerable. These are to be reported to the authors.  

Moreover, we found a function named~\texttt{retry(address)} in the dataset that is marked as vulnerable for RE even though the function is non-vulnerable\footnote{\href{https://github.com/InPlusLab/ReentrancyStudy-Data/blob/main/reentrant_contracts/0x7c4393ee129d7856b5bd765c2d20b66f464ccd0f.sol}{SC-link}}. It is protected by the modifier that restricts function calling i.e. \texttt{onlyWhitelisted(address)}.

\subsection{Manually Injected (MI) Dataset}

Ren et al.~\cite{Empirical-Evaluation-of-Smart-Contract-Testing:What-is-the-Best-Choice} created two labelled datasets. One dataset is called Manually Injected (MI) bugs, created by injecting bugs into different locations of SCs. MI dataset has seven different types of vulnerabilities with 50 contracts per vulnerability. The paper mentions that the logic for vulnerabilities is simple and involves only a few obvious patterns. Hence, this dataset does not cover all RE and UX scenarios. Another issue with the MI dataset is the absence of non-vulnerable functions, which makes it impossible to check the false positives of existing tools. The second dataset, the CVE dataset, does not include RE or UX vulnerabilities and, therefore, is not considered when discussing limitations.


\subsection{Majority Voting Dataset}

\begin{table}[h!]
	\centering
	\begin{tabular}{|l|l|l|}
		\hline
		\textbf{Tool}     & \textbf{RE} & \textbf{UX} \\ \hline
		Oyente\cite{oyente}            & \checkmark & \crossmark                                  \\ \hline
		Mythril\cite{mythril}           & \checkmark & \checkmark                                 \\ \hline
		Securify\cite{securify}          & \checkmark & \checkmark                                 \\ \hline
		Smartcheck\cite{smartcheck}        &  \crossmark & \checkmark                                 \\ \hline
		Pied-Piper\cite{pied-piper}       & \crossmark & \crossmark                                  \\ \hline
	\end{tabular}
	\caption{RE and UX Vulnerabilities Supported by Tools in the Majority Voting Framework by \cite{Making-Smart-Contract-Development-More-Secure-and-Easier}}
	\label{tab:majority-voting-tool-support}
\end{table}

Table~\ref{tab:majority-voting-tool-support} displays the tools employed by Ren et al.~\cite{Making-Smart-Contract-Development-More-Secure-and-Easier}, noting that outdated tools such as Oyente and Securify are included. Among these tools, only three support RE and UX vulnerabilities, with Oyente and Securify known to produce false positives\cite{Empirical-vulnerability-analysis-of-automated-smart-contracts-security-testing-on-blockchains, Empirical-Review-of-Automated-Analysis-Tools-on-47587-Ethereum-Smart-Contracts}. The framework proposed by Ren et al. employs a majority voting system, where a contract is classified as vulnerable if at least two out of three tools concur on the presence of a vulnerability, specifically for RE or UX vulnerabilities. However, the reliability of this approach is compromised by the high rate of false positives reported by two of the tools. This can lead to erroneous classifications of contracts as vulnerable, even when they are not, thereby reducing the effectiveness of the framework in accurately identifying genuine vulnerabilities.

\subsection{Smartbugs Dataset}

Durieux et al.~\cite{Empirical-Review-of-Automated-Analysis-Tools-on-47587-Ethereum-Smart-Contracts} crafted an annotated and non-annotated SCs dataset. The annotated part contains~\totalLabelledRENTSCinSmartbugs~contracts tagged with RE vulnerability. Total functions in these contracts are~\totalLabelledRENTFuninSmartbugs, out of which only~\totalLabelledBuggyFuninSmartbugs~are vulnerable for RE. Moreover, we found a function named~\texttt{WithdrawToHolder(address,uint)} in the dataset that is marked as vulnerable for RE even though the function is non-vulnerable\footnote{\href{https://github.com/smartbugs/smartbugs-curated/blob/main/dataset/reentrancy/0x627fa62ccbb1c1b04ffaecd72a53e37fc0e17839.sol}{SC-link}}. It is protected by the modifier that restricts function calling i.e. \texttt{onlyOwner}~\cite{onlyowner}. A similar concept is demonstrated in Listing~\ref{lst:reent-struct}, in the function~\texttt{non\_vuln\_onlyOwner()}.

\section{\mytoolcd: SCs Labelled using Crowdsourcing}
\label{sec:crowdsource}
This section explains the methodology for creating a crowdsource dataset. 


\subsection{Data Source}
We now explain the collection methodology for contracts in~\mytoolcd.
\begin{figure}[ht]
    \hspace*{0.5cm} 
  \resizebox{0.9\textwidth}{!}{%
\begin{lstlisting}[
      captionpos=b, 
      escapechar=\%, 
      language=Solidity, 
      label=lst:bigquery_tx_gt_1, 
      frame=single, 
      caption={Query to Retrieve SC Addresses with Transaction Counter Greater Than One from Google BigQuery}
    ]
SELECT contracts.address, balances.eth_balance total_eth_balance, COUNT(transactions.to_address) AS transaction_count 
FROM `bigquery-public-data.crypto_ethereum.contracts` AS contracts 
JOIN `bigquery-public-data.crypto_ethereum.balances` AS balances ON balances.address = contracts.address 
LEFT JOIN `bigquery-public-data.crypto_ethereum.transactions` AS transactions ON transactions.to_address = contracts.address 
GROUP BY contracts.address, total_eth_balance 
HAVING transaction_count > 1 
ORDER BY total_eth_balance DESC;
\end{lstlisting}
  }
\end{figure}

We utilized the \textit{crypto\_ethereum} table from Google BigQuery~\cite{bigquery}, a leading platform that provides access to diverse organizational databases.  The \textit{crypto\_ethereum} table from Google BigQuery serves as a comprehensive repository for the most up-to-date Ethereum SC addresses. To filter out dummy SCs we consider SCs with transaction count greater than one. Running the query in Listing~\ref{lst:bigquery_tx_gt_1} on Google BigQuery yielded~\totalSCtxGTOne~SC addresses.


Since Google BigQuery does not provide source codes, we used the Etherscan API\footnote{\href{https://docs.etherscan.io/api-endpoints/contracts\#get-contract-source-code-for-verified-contract-source-codes}{https://docs.etherscan.io/api-endpoints/contracts\#get-contract-source-code-for-verified-contract-source-codes}} to retrieve the contract source code, extracting~\extractedSCtxGTOne~source codes out of~\totalSCtxGTOne~contracts, as many SCs do not make their source code public and due to Etherscan API extraction limitations. We performed de-duplication, following the approach adopted from~\cite{Empirical-Review-of-Automated-Analysis-Tools-on-47587-Ethereum-Smart-Contracts}, to obtain unique smart contracts (SCs). This process involved removing newlines, spaces, and comments from the SCs, followed by exactly matching the cleaned versions to identify and eliminate duplicates. After performing de-duplication on extracted SCs, we got~\uniqueExtractedSCtxGTOne~unique SCs. We labelled~\totalSCInCrowd~contracts, chosen for their mix of simple to complex functions and coverage across different Solidity versions. Our rigorous procedure limited the number we could annotate. We plan to continue the labelling process to improve the dataset. 


\subsection{Tools Used}
\label{sec:crowd-tools-used}
As in previous work~\cite{Empirical-Review-of-Automated-Analysis-Tools-on-47587-Ethereum-Smart-Contracts}, we adopt the following criteria for selecting the tools: 
\begin{enumerate}
	\item The tool is publicly released and provides a command-line interface (CLI). 
	\item The tool accepts Solidity\footnote{Solidity is widely used programming language used to write SCs.} SCs as input. 
	\item The tool should be scalable to large programs, actively maintained, and provide clear setup instructions. 
\end{enumerate}


We selected Slither\cite{slither}, Mythril\cite{mythril}, Solhint\cite{solhint}, Conkas\cite{conkas}, and Sailfish\cite{sailfish} for our analysis, while excluding dynamic tools such as ityfuzz\cite{shou2023ityfuzz}, Reguard\cite{reguard}, SoliAudit\cite{soliaudit-ml}, ConFuzzius\cite{ConFuzzius}, and ContractFuzzer\cite{contractfuzzer} due to their scalability limitations. Additionally, we excluded Oyente\cite{oyente} and Securify\cite{securify} for being outdated.

\subsection{\mytoolcd~Methodology}
\label{sec:crowd-method}
\begin{figure}[ht]
  \centering
  \fbox{%
    \includegraphics[width=0.95\columnwidth]{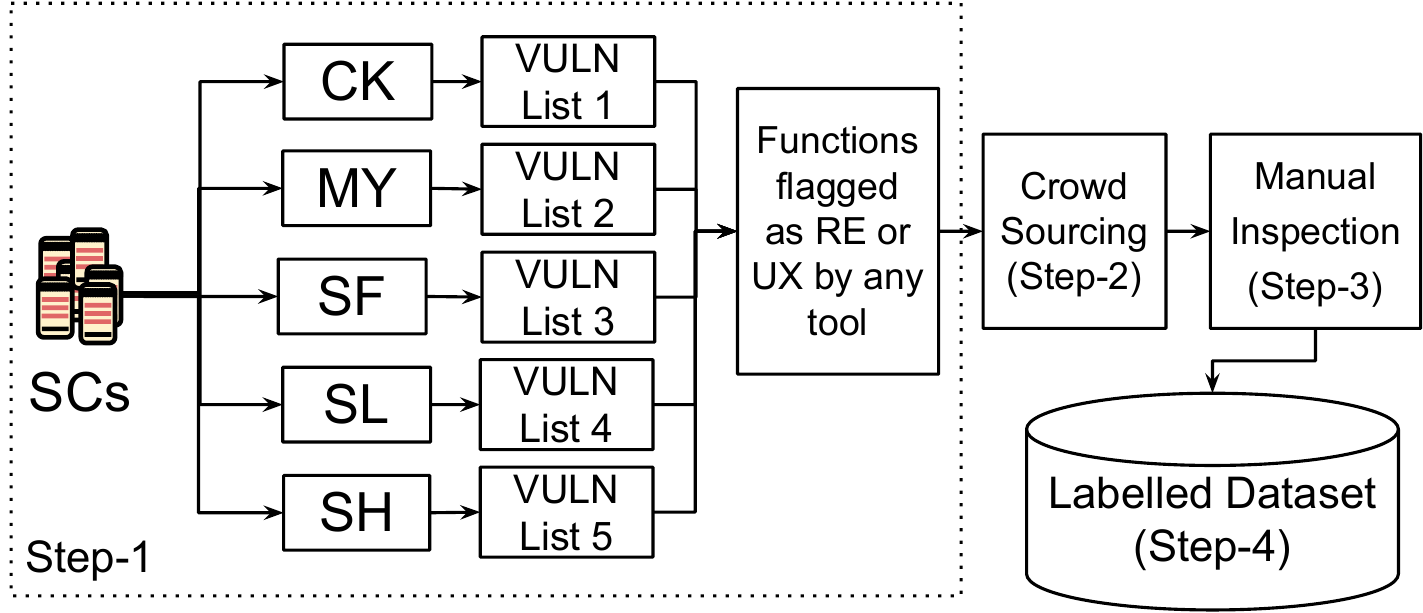}
  }
  \caption{Methodology to create~\mytoolcd~(Conkas: CK, Mythril: MY, Sailfish: SF, Slither: SL, Solhint: SH)}
  \label{fig:crowdsource}
\end{figure}

Figure~\ref{fig:crowdsource} illustrates our rigorous methodology for creating~\mytoolcd. The goal of employing the crowdsourcing approach is to gather input from graduate-level Computer Science and Engineering (CSE) students to label real-world SCs. The steps are as follows:
\begin{enumerate}
    \item We initially taught students the basics of blockchain, smart contracts, and vulnerabilities in a graduate-level course in the Department of CSE (Step-0).
    \item  The SC vulnerability task was assigned via a website, where functions flagged as RE or UX by at least one tool (Solhint, Sailfish, Mythril, Conkas, Slither) were reviewed by students. Each student analyzed 20 functions, with a maximum of three students per function (Step-1).
    \item  We gathered input from~\totalStudentsinCrowdsource~students (Step-2). The annotations were reviewed and validated by one of the authors, an expert SC programmer (Step-3).
    \item The conflicting labels were finalized after one-to-one meeting with students. As a result, we have dataset with high confidence. Moreover, we have proper comments explaining why a vulnerability exists or does not exist for non-trivial functions (Step-4). 
\end{enumerate}
   Understanding a SC function is an intense task and our crowdsourcing methodology is quite rigorous. Therefore we cannot expect humans to be consistent if a large number of SC functions are given for annotation. Hence, we decided to give \textit{20} functions to each student. The total number of functions flagged as vulnerable for RE or UX by at least one tool is~\totalFunsMarkedVulnByTools~from~\totalSCInCrowd~contracts. Due to the reasons specified above, we were able to annotate~\totalRENTFunsInCrowd~functions for RE and~\totalUEFunsInCrowd~for UX from~\totalSCInCrowd~contracts. Out of~\totalRENTFunsInCrowd~functions,~\totalRENTVulnFunsInCrowd~are vulnerable to RE while the rest~\totalRENTNonVulnFunsInCrowd~are non-vulnerable. Out of the~\totalRENTFunsInCrowd~functions for RE,~\totalRENTStudentFunsInCrowd~were annotated with the help of students, while~\totalRENTNonStudentFunsInCrowd~were annotated directly by an experienced SC programmer. Within the same~\totalSCInCrowd~contracts, we have~\totalUEFunsInCrowd~functions labelled for UX Vulnerability. Out of~\totalUEFunsInCrowd~functions,~\totalUEVulnFunsInCrowd~are vulnerable to UX, whereas the rest~\totalUENonVulnFunsInCrowd~are non-vulnerable. All UX functions were annotated with the help of students (see Table~\ref{table:vulnerability-overview}).
  


\section{\mytoolsd: Manually Synthesized Reentrancy Dataset}

\mytoolsd~covers corner cases of RE vulnerability which are not present in~\mytoolcd. It includes test cases for both vulnerable and non-vulnerable scenarios, where a minor modification can switch a test case between vulnerable and non-vulnerable. 


    
\subsection{Methodology}
\label{sec:man-method}
Two of the authors are expert SC programmers and have studied RE vulnerability in-depth. Using their experience, they manually created various scenarios that result in RE and other scenarios that are very similar but do not result in RE. We have compiled these scenarios into a dataset focused on RE vulnerabilities, which covers diverse cases, such as data dependencies, control dependencies, RE vulnerabilities through require/assert statements, and multi-call issues. We believe this dataset encompasses nearly all potential RE scenarios. \mytoolsd~contains a total of~\totalFunsinManualDB~functions, out of which~\totalRENTVulnFunsInManual~are vulnerable and~\totalRENTNonVulnFunsInManual~are non-vulnerable. Table~\ref{table:vulnerability-overview} describes the result of annotations. ~\mytoolsd~can be used as a test suite to check if the existing tools cover complex as well as simple RE scenarios.~\mytoolsd~currently does not handle UX.

\begin{figure*}[ht]

	\hspace*{0.5cm} 
			\centering
		\resizebox{\textwidth}{!}{%
\begin{lstlisting}[
			captionpos=b, 
			escapechar=\%, 
			language=Solidity, 
			label=lst:reent-struct, 
			frame=single, 		
			caption={Example RE Scenarios in \mytoolsd}
			]
// SPDX-License-Identifier: MIT
pragma solidity ^0.8.0;

contract ReentrancyExampleContract {
    address public owner;
    uint256 public state_var;
    uint256 public state_var_1;
    uint256 public state_var_2;
    // Modifier to restrict access to only the owner
    modifier onlyOwner() { 
        require(msg.sender == owner, "Only the owner can execute this.");
        _;
    }
    constructor() {
        owner = msg.sender;
        state_var = 1;
        state_var_1 = 8;
        state_var_2 = 2;
    }
    function vuln_data_dep() public {  // parameter of call is data detependent on st updated post calpubll
        // Initially , state_var_1 = 1
        msg.sender.call{value: state_var}("");
        state_var = state_var - 1; 
    }
    // Vulnerable function where execution depends on a condition and state change
    function vuln_control_dep() public {
        // Initially , state_var_1 = 1
        if (state_var_1 > 0)
            msg.sender.call{value: state_var_2}(""); // Executes only if the condition is true
        state_var_1 = state_var_1 - 1; // State variable is updated after the call
    }
    // Non-vulnerable function with onlyOwner modifier
    function non_vuln_onlyOwner() public onlyOwner {
        msg.sender.call{value: state_var}(""); // Only owner can call this
        state_var = state_var - 1; // State variable is updated after the call
    }
    // Vulnerable multi-call function that can allow reentrancy
    function vuln_multi_call(address address_1, address address_2) public {
        address_1.call{value: address(this).balance / 2}(""); // address_1 can reenter multiple times until all Ether gets drained
        address_2.call{value: address(this).balance / 2}(""); // address_2 won't get any Ether if address_1 reenters until all Ether gets drained
    }
    // Vulnerable function where require/assert can be manipulated by reentrancy
    function vuln_require_assert() public {
        // Initially , state_var_1 = 8
        require(state_var_1 < 10, "State variable is not less than 10"); // OR assert(state_var_1 < 10);
        msg.sender.call{value: state_var_2}(""); // If attacker reenters from here, state_var_1 will remain 8 and require condition will be satisfied
        state_var_1 = state_var_1 + 4; // If executed, state_var_1 will be 12 and condition will fail
    }
}
\end{lstlisting}
		}
\end{figure*}

    \subsection{Reentrancy Scenarios}
    \label{sec:man-rent-scen}
    \subsubsection{RE Induced by Data Dependency}
	The function \texttt{vuln\_data\_dep()} in Listing~\ref{lst:reent-struct} illustrates a data dependency scenario, where the external call's parameter (\texttt{state\_var}) is updated after the call. If an attacker reenters before the update, they can receive Ether they should not get. Thus, \textit{the external call's parameter depends on the state variable updated after the call.} Consider the initial value of \texttt{state\_var} as 1. If the function executes sequentially, \texttt{state\_var} will decrement to zero after the first call, preventing any ether from being sent in subsequent calls. However, if an attacker exploits reentrancy by making an external call, they can bypass this decrement and repeatedly withdraw 1 wei~\footnote{wei is a unit of Ether} each time. This illustrates how reentrancy can manipulate contract behavior to siphon funds.
	
    \subsubsection{RE Induced by Control Dependency}
    The function \texttt{vuln\_control\_dep()} from Listing~\ref{lst:reent-struct} shows a function in which \textit{an external call is control dependent on state variable updated after the external call (\texttt{state\_var\_1}).} Here, an attacker can reenter the function before the updation of an external call, and the condition will always be true. Meanwhile, the condition can become false in the non-reentrant scenario after updating the state variable. Initially, \texttt{state\_var\_1} is set to 1, allowing the external call to proceed as the condition (\texttt{if (state\_var\_1 > 0)}) holds true. If the function \texttt{vuln\_control\_dep()} executes sequentially, \texttt{state\_var\_1} is decremented to zero, rendering the condition false in any subsequent calls. However, if an attacker reenters the function during the external call (via \texttt{msg.sender.call\{value: state\_var\_2}("")\}), the conditional check will not prevent execution, enabling the attacker to continuously extract ether. This scenario demonstrates how reentrancy can exploit control flow conditions to siphon funds from the contract.
    \subsubsection{RE Induced by Modifier that Restricts Function Calling}
    The function \texttt{non\_vuln\_onlyOwner()} from Listing~\ref{lst:reent-struct} shows an example where the function is protected by \texttt{onlyOwner()} modifier~\cite{onlyowner}. Here, the state variable \texttt{owner} denotes the address of the owner of the contract. The modifier ensures that only the owner can call the function. An attacker cannot reenter such a function because of protection from the modifier. 
    \subsubsection{RE Induced by Multi Call}
    The function \texttt{vuln\_multi\_call()} from Listing~\ref{lst:reent-struct} shows an example of reentrancy vulnerability due to multiple external calls that belong to different addresses. In case of sequential execution \texttt{address\_1} will get half ether and \texttt{address\_2} will get half ether.  If address \texttt{address\_1} is malicious then it can reenter the function till it drains all Ether from the contract. Address \texttt{address\_2} won't get any Ether. 
    \subsubsection{RE Induced by Control Dependency due to require() and assert()}
    The function \texttt{vuln\_require\_assert()} shown in Listing~\ref{lst:reent-struct} demonstrates a control dependency scenario stemming from a \texttt{require()} statement. It's important to note that replacing \texttt{require()} with \texttt{assert()} in this function would still retain the vulnerability. Initially, consider \texttt{state\_var\_1} set at 8. With the condition \texttt{require(state\_var\_1 < 10)} satisfied, the external call is executed. If the function is executed sequentially, \texttt{state\_var\_1} is incremented by 4 to 12, making the \texttt{require()} condition false in subsequent calls, which halts further execution. However, if an attacker exploits reentrancy during the external call, the \texttt{require} condition still holds, potentially allowing them to drain all ether from the smart contract. This demonstrates how reentrancy can enable unintended ether withdrawals.
    
\section{Evaluation of Existing Tools using~\mytool}
\label{sec:results}


\begin{figure}[t]
  \centering
  \begin{subfigure}{0.23\textwidth}
    \centering
    \includegraphics[width=\linewidth]{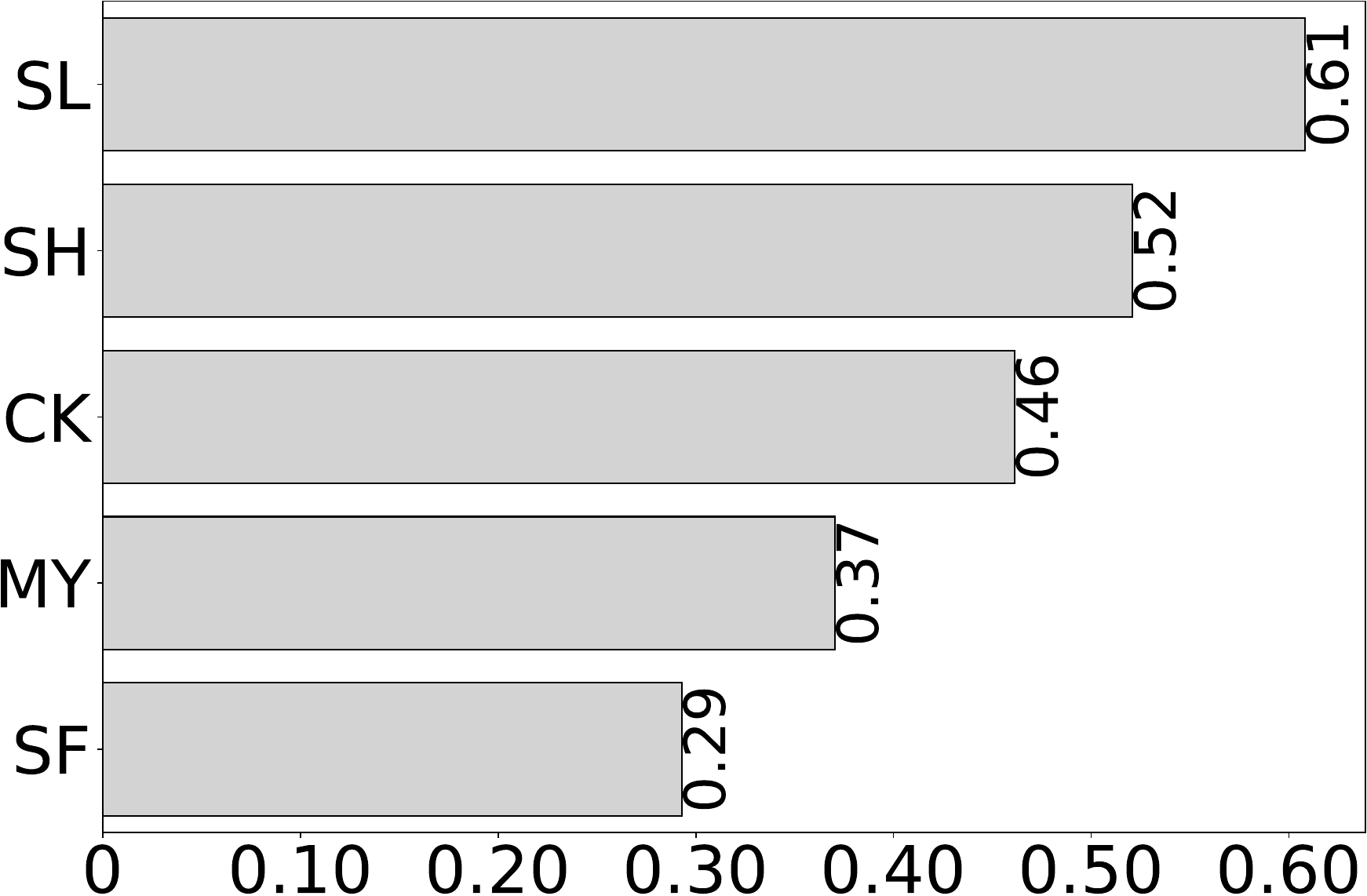}
    \caption{\mytoolcd~(RE)}
    \label{fig:f1-crowd-rent}
  \end{subfigure}%
  \hspace{0.02\textwidth} 
  \begin{subfigure}{0.23\textwidth}
    \centering
    \includegraphics[width=\linewidth]{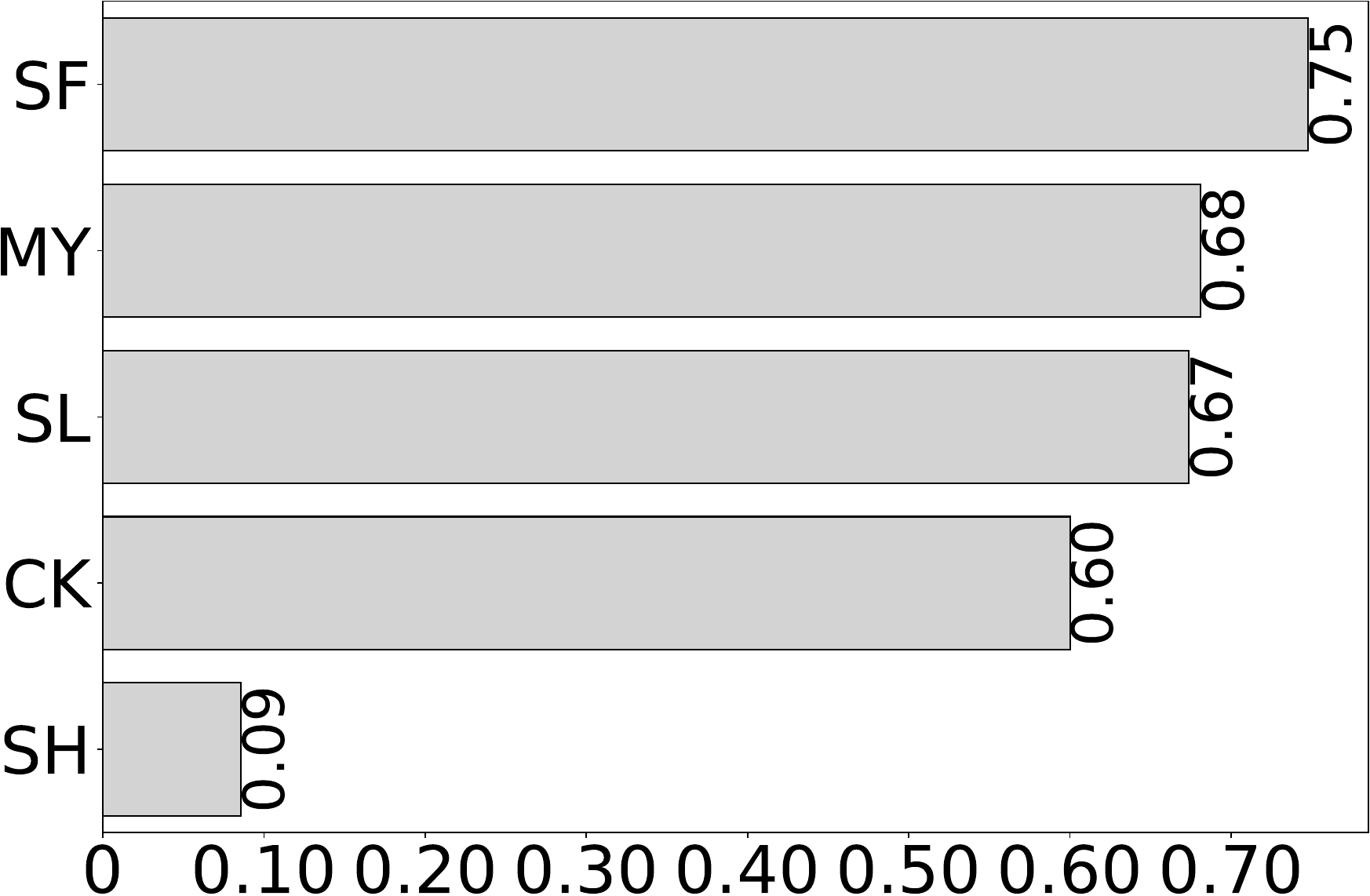}
    \caption{\mytoolsd~(RE)}
    \label{fig:f1-manual-rent}
  \end{subfigure}
  \begin{subfigure}{0.23\textwidth}
    \centering
    \includegraphics[width=\linewidth]{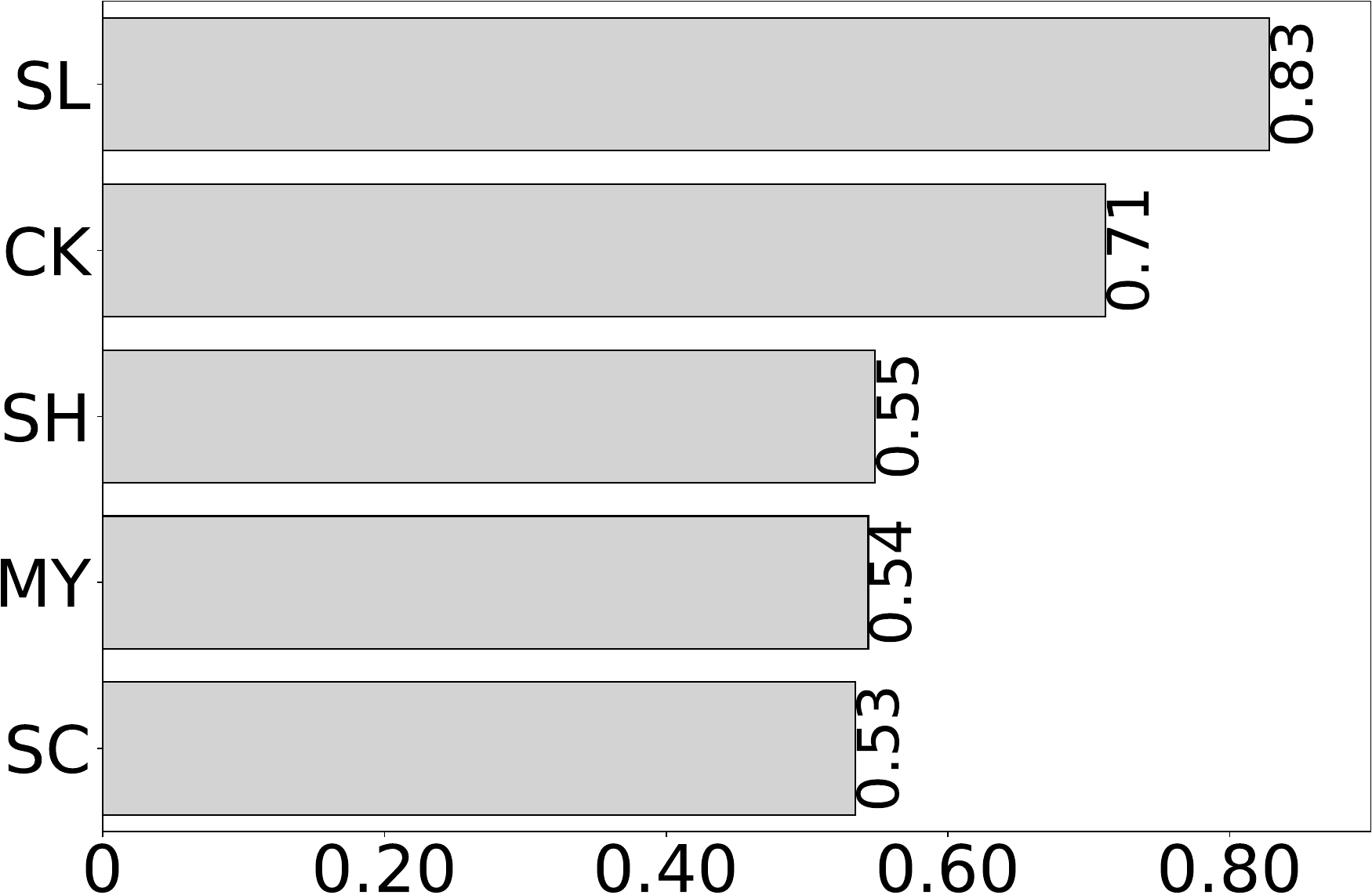}
    \caption{\mytoolcd~(UX)}
    \label{fig:f1-crowd-ue}
  \end{subfigure}
  \caption{F1-Scores of Existing Tools using~\mytool ~(Slither: SL, Sailfish: SF, Solhint: SH, Mythril: MY, Conkas: CK, Smartcheck: SC) — Higher the Better}
\end{figure}

\begin{table}[h!]
	\centering
	\begin{tabular}{|l|l|l|}
		\hline
		\textbf{Tool}     & \textbf{RE} & \textbf{UX} \\ \hline
		Mythril\cite{mythril} & \checkmark & \checkmark                                 \\ \hline
		Slither\cite{slither} & \checkmark & \checkmark                                 \\ \hline
		Solhint\cite{solhint} & \checkmark & \checkmark                                 \\ \hline
		Conkas\cite{conkas} & \checkmark & \checkmark                                 \\ \hline
		Smartcheck\cite{smartcheck}  &  \crossmark & \checkmark                                 \\ \hline
		Sailfish\cite{pied-piper}    & \checkmark & \crossmark                                  \\ \hline
		
	\end{tabular}
	\caption{RE and UX Vulnerabilities Supported by Tools used for Evaluation}
	\label{tab:tools-used-evaluation}
\end{table}

This section shows F1-Score of existing tools (Slither~\cite{slither}, Mythril~\cite{mythril}, Sailfish~\cite{sailfish}, Conkas~\cite{conkas}, Smartcheck~\cite{smartcheck}, Solhint~\cite{solhint}) while detecting RE and UX vulnerabilities from~\mytool. Table~\ref{tab:tools-used-evaluation} shows the support provided by these tools for RE and UX vulnerabilities.



Figure~\ref{fig:f1-crowd-rent} shows the F1-Score of existing tools while detecting RE on~\mytoolcd. The F1-Score is a metric used to evaluate the performance of a classification model, and it is the harmonic mean of precision and recall~\cite{f1score}. Slither (SL) outperforms other tools because it over-approximates RE. \mytoolcd~do not cover all corner cases of RE. Figure~\ref{fig:f1-manual-rent} shows the F1-Score of existing tools on~\mytoolsd. Here, Sailfish outperforms other tools because it handles corner cases of RE present in~\mytoolsd. Smartcheck~\cite{smartcheck} does not support detection of RE and hence omitted.  

Figure~\ref{fig:f1-crowd-ue} shows the performance of existing tools on UX vulnerabilities from~\mytoolcd. Sailfish does not support the detection of UX and hence omitted. Here, we can see that Slither outperforms other tools in detecting UX vulnerabilities. Sailfish~\cite{sailfish} does not support detection of UX and hence omitted. 
\section{Applications of~\mytool}
\mytool~has a total of~\totalRENTFunsInScrawlD~functions labelled with RE vulnerability and~\totalUEFunsInCrowd~functions labelled with UX vulnerability. \mytool~captures complex RE and UX scenarios. \mytool~dataset includes non-vulnerable scenarios that are incorrectly flagged as vulnerable by existing tools. Moreover,~\mytoolcd~captures real world scenarios of unhandled exception vulnerability. Given this information, following subsections point out the applications of~\mytool.
    \begin{enumerate}
    \item \textbf{Evaluation of New and Existing Tools: }
    Researchers can check if existing as well as new tools cover the complex scenarios of Reentrancy Vulnerability and Unhandled Exceptions vulnerability using~\mytool.
    \item \textbf{Training of Machine Learning based Tools: }
    Machine learning-based SC vulnerability analysis tools~\cite{mlsc1, mlsc2, mlsc3} require a labelled dataset to be trained.~\mytool~can serve as a foundation for researchers to learn or train machine learning-based tools.

    \end{enumerate}

\section{Conclusion and Future Work}


In this work, we introduced~\mytool, a dataset designed to capture a wide range of RE and UX vulnerabilities in smart contracts. The dataset consists of two parts:~\mytoolcd~and~\mytoolsd. The~\mytoolcd~portion contains real-world smart contracts labeled for RE and UX vulnerabilities, created using crowdsourcing, while~\mytoolsd~was developed by carefully crafting additional RE scenarios not covered in~\mytoolcd. The dataset includes~\totalRENTFunsInScrawlD~functions labeled with RE vulnerability and~\totalUEFunsInCrowd~functions labeled with UX vulnerability. We evaluated several existing vulnerability analysis tools on~\mytool~and~found that Slither outperforms other tools in detecting RE and UX vulnerabilities in contracts from~\mytoolcd, while Sailfish excels in detecting RE vulnerabilities from~\mytoolsd. Going forward, we plan to conduct a more detailed empirical evaluation of these tools using~\mytool. Additionally, we aim to explore the use of machine learning techniques to automatically annotate the remaining contracts in~\mytool.

\bibliographystyle{IEEEtran}
\bibliography{sections/references}
\end{document}